\documentstyle[12pt]{article}
\newcommand{\be}{\begin{equation}}
\newcommand{\en}{\end{equation}}
\newcommand{\bea}{\begin{eqnarray}}
\newcommand{\ena}{\end{eqnarray}}
\newcommand{\Det}{\hbox{Det}}
\newcommand{\hbo}{\hbox to 1 true cm {\hfill } }
\newcommand{\tr}{\hbox{tr}}

\def\dslash{\partial\kern-.5em\slash}
\def\kslash{k\kern-.5em\slash}
\def\pslash{p\kern-.5em\slash}

\begin{document}
\vglue 1truecm

\vbox{ T95/072
\hfill June 14, 1995
}
\vbox{ hep-ph/9506341 \hfill }

\vfil
\centerline{\bf Chiral symmetry breaking in strongly coupled scalar QED$^1$}

\bigskip
\centerline{ M.\ Chemtob, K.\ Langfeld }
\vspace{1 true cm}
\centerline{ Service de Physique Theorique, C.E.A. Saclay, }
\centerline{F--91191 Gif--sur--Yvette Cedex, France. }
\bigskip

\vfil
\begin{abstract}

Scalar QED with fermions is investigated in an
expansion in powers of the inverse gauge coupling constant.
The fermion mass generation is studied in next to leading order
of the strong coupling expansion for the Higgs-phase of the model.
Chiral symmetry breaking is discussed. Our approach supports from
a strong coupling point of view the results obtained earlier by Miransky,
Bardeen and Kondo.

\end{abstract}

\vfil
\hrule width 5truecm
\vskip .2truecm
\begin{quote}
$^1$ Supported by DFG under contracts La--$932/1-1/2$.
\end{quote}
\eject
\section{ Introduction }
\label{sec:1}

The strong coupling behaviour of U(1) gauge theories interacting
with scalar and/or fermionic matter is a largely unsolved problem.
Quantum electrodynamics (QED) is the prototype of non-asymptotically
free or slowly running (walking) gauge theories, which could
provide useful insight into the important issue of chiral symmetry
breaking~\cite{nam61}. Intensive discussions in the late 80's have
suggested the intriguing possibility that
non-asymptotically free gauge theories could exhibit a second order
phase transition to the chiral broken phase at large values of the
coupling constants
$e^2 \ge e_c^2= 4 \pi^2/3$. In the pioneering work by Miransky at
al.~\cite{mir85}, it was first pointed out that
the renormalisation procedure imposes a specific cutoff dependence on the
bare coupling strength, which yields a peculiar scaling law of the
essential-singularity type for the running coupling strength~\cite{mir85}.
However, such a scaling of the coupling strength is inconsistent, because
this result was derived in quenched approximation where the electric charge
needs no renormalisation at all. This led Bardeen, Leung and Love~\cite{bar86}
to include a Nambu-Jona-Lasinio type four fermion interaction.
They argued that this interaction must be taken into account,
since it becomes relevant in the continuum limit due to a large
non-perturbative anomalous dimension. Subsequently the  critical exponents
of the renormalisation
group flow were extracted from the analysis of the Dyson-Schwinger
equation~\cite{boo89} and from the lattice formulation~\cite{koc90}.
The fact that the critical coupling constant $e_c$ lies
in any case above unity makes it desirable to develop
an alternative formulation of QED using a perturbative expansion
in inverse powers of the gauge coupling constant.

In this paper, we propose a new formulation of scalar QED
(termed as dual formulation) in the Higgs phase coupled to fermions,
which leads to the strong coupling expansion in the electric charge,
and will provide further
support for the ideas presented earlier~\cite{mir85,bar86,co89,ko89,kon91}.
We will find that an effective chirally symmetric four fermion
interaction of the type introduced by Bardeen, Leung and
Love~\cite{bar86} naturally arises in the strong coupling expansion.
We will therefore provide evidence, that the results obtained earlier
might be valid even for the large value of the coupling strength
of interest ($e^2 \approx  4\pi ^2/3$).

In order to address the question of spontaneous symmetry breaking in
our approach, we will search for a non-trivial solution of the
Dyson-Schwinger equation for the fermionic self-energy.
Our studies differ from that of Kondo~\cite{ko89}, who considered the case
of a massive photon, by including the Nambu-Jona-Lasinio contact
interaction, and from that of Bardeen et al.~by taking into account
the mass of the photon.
We will find that chiral symmetry breaking
occurs also for small values of the gauge coupling constant due
to the support from the contact interaction. This is in agreement
with the conclusions of~\cite{bar86}.

The paper is divided in two parts. In the first part (section 2)
the dual formulation is derived, gauge fixing is discussed and the
effective fermion interaction is obtained consistent with the second
order of the strong coupling expansion.
In the second part (section 3) the Dyson-Schwinger equation for the
fermion self-energy is studied using the obtained fermion interaction.
Chiral symmetry breaking is addressed.
Conclusions are summarised in the final section.

\section{ The dual formulation and the strong coupling expansion }
\subsection{ Reformulating scalar QED }

The Euclidean generating functional for photon insertions of scalar
QED is given by a functional integral over the complex scalar field
$\phi $ and the photon field $A_\mu $, i.e.
\bea
Z[j] &=& \int {\cal D} \phi \; {\cal D} \phi ^{\dagger } \;
{\cal D} A_\mu \; \exp \left\{ - \int d^{4}x \; \left[
{\cal L } \, - \, j_\mu (x) A_\mu (x) \right] \right\} \; ,
\label{eq:1} \\
{\cal L } &=& \left( D_\mu \phi \right)^\dagger D_\mu \phi
\; + \; \frac{1}{ 4 e^2 } F_{\mu \nu }[A] F_{\mu \nu }[A]
\; + \; U \left( \vert \phi \vert ^2 \right) \; ,
\label{eq:2}
\ena
where the field strength functional is
\be
F_{\mu \nu }[A] \; = \; \partial _\mu A_\nu (x) \, - \, \partial _\nu
A_\mu(x) \; ,
\label{eq:3}
\en
and $U$ serves as a potential for the scalar fields.
The gauge covariant derivative $D_\mu := \partial _\mu + i A_\mu (x)$
is chosen to transform homogeneously under U(1) gauge transformation, i.e.
\be
\phi (x) \rightarrow \phi '(x)= e^{i \alpha (x) } \phi (x) \; , \hbo
A_\mu(x) \rightarrow A_\mu '(x)=A_\mu (x) - \partial _\mu \alpha (x) \; .
\label{eq:4}
\en
In order to derive the dual formulation of the theory defined by (\ref{eq:1})
we follow the ideas of the field strength approach to non-abelian
Yang-Mills theories~\cite{ha77} and
introduce an anti-symmetric tensor field $T_{\mu \nu }$ and a vector
field $V_\mu $, which are both singlets under gauge transformation, i.e.
\bea
Z[j] &=& \int {\cal D} \phi \; {\cal D} \phi ^{\dagger } \;
{\cal D} A_\mu \; {\cal D} T_{\mu \nu } \; {\cal D} V_\mu \;
\exp \left\{ - \int d^{4}x \; \left[
{\cal L }_E \, - \, j_\mu (x) A_\mu (x) \right] \right\} \; ,
\label{eq:5} \\
{\cal L }_E &=& \left( \partial _\mu \phi \right)^\dagger
\partial _\mu \phi \, + \, \vert \phi \vert ^2 A_\mu A_\mu
\, - \, A_\mu J_\mu
\label{eq:6} \\
&+& \frac{ e^2 }{4} T_{\mu \nu } T_{\mu \nu } \, + \, \frac{i}{2}
T_{\mu \nu } F_{\mu \nu }[A] \, + \, i V_\mu \partial _\nu
\widetilde{T} _{\mu \nu } \; + \; U \left( \vert \phi \vert ^2 \right) \; ,
\nonumber
\ena
where $J_\mu $ is the electric current of the scalar field (without
the diamagnetic term), i.e.
\be
J_\mu (x) := i \left( \phi ^{\dagger } \partial _\mu \phi
\; - \; \partial _\mu \phi ^{\dagger } \phi \right) \; ,
\label{eq:7}
\en
and the tilde denotes the dual field strength, e.g.\
$\widetilde{T}_{\mu \nu } \; = \; \frac{1}{2} \epsilon _{\mu \nu \alpha \beta }
T_{\alpha \beta }$. It is easy to check that the generating functional
in (\ref{eq:5}) is equivalent to that in (\ref{eq:1}). For this purpose
we integrate out the tensor field $T_{\mu \nu }$ in (\ref{eq:5}).
The resulting effective theory of the fields $\phi $, $A_\mu $ and $V_\mu $
is described by the Lagrangian
\bea
L_{eff} &=& \left( \partial _\mu \phi \right)^\dagger
\partial _\mu \phi \, + \, \vert \phi \vert ^2 A_\mu A_\mu
\, - \, A_\mu (j_\mu + J_\mu ) \, + \, \frac{1}{4 e^2} F_{\mu \nu }
F_{\mu \nu }
\label{eq:8} \\
&-& \frac{1}{e^2} \partial _\mu \widetilde{F}_{\mu \nu }[A] V_\nu
\, + \, \frac{1}{e^2} \partial _\mu V_\nu \partial _\mu V_\nu \; + \;
U \left( \vert \phi \vert ^2 \right) \; .
\label{eq:9}
\ena
The first line (\ref{eq:8}) is precisely the Lagrangian of scalar QED
(\ref{eq:2}). The second line (\ref{eq:9}) contains the interaction with
the vector field $V_\mu $. The crucial observation is that the
vector field $V_\mu $ decouples due to Bianchi's identity, i.e.\
$\partial _\mu \widetilde{F}_{\mu \nu }[A] = 0$, and the fact that the
integration
over the vector field yields an unimportant constant. The basic idea to
introduce the vector field $V_\mu $ is to reduce the
degrees of freedom of the formulation (\ref{eq:5}) to the physical ones
as we will see below.

Since we have established the equivalence of both formulations of
scalar QED (\ref{eq:1}) and (\ref{eq:5}), we proceed further with
the formulation (\ref{eq:5}) by integrating out the photon field
$A_\mu $ and the vector field $V_\mu $. The generating functional is
\bea
Z[j] &=& \int {\cal D} \phi \; {\cal D} \phi ^{\dagger } \;
{\cal D} T_{\mu \nu } \; \delta \left( \partial _\nu
\widetilde{T} _{\mu \nu } \right) \; \Det ^{-2} \vert \phi \vert ^2 \;
\exp \left\{ - \int d^{4}x \; {\cal L }_T \right\} \; ,
\label{eq:10} \\
{\cal L }_T &=& \left( \partial _\mu \phi \right)^\dagger
\partial _\mu \phi \, + \, \frac{1}{ 4 \vert \phi \vert ^2 }
\left( \partial _{\nu } T_{\nu \mu } -i j_\mu - i J_\mu \right)^2
\, + \, \frac{ e^{2} }{4} T_{\mu \nu } T_{\mu \nu } \; + \;
U \left( \vert \phi \vert ^2 \right) \; .
\label{eq:11}
\ena
Integrating out $V_\mu $ in (\ref{eq:5}) constrains the tensor field
$T_{\mu \nu }$ by Bianchi's identity, i.e.\ $ \partial _\nu
\widetilde{T} _{\mu \nu } =0 $. In order to solve this constraint we
decompose the tensor field, i.e.
\be
T_{\mu \nu } \; = \; \partial _\mu C_\nu - \partial _\nu C_\mu \; + \;
\epsilon _{\mu \nu \alpha \beta } \partial _\alpha a_\beta \; .
\label{eq:12}
\en
The tensor field $T_{\mu \nu }$ has six degrees of freedom, whereas
the fields $C_\mu $, $a_\mu $ count eight. However, two degrees are
redundant, since the gauge transformations
\be
C_\mu(x) \rightarrow C_\mu '(x)=C_\mu (x) - \partial _\mu \sigma (x) \; , \hbo
a_\mu(x) \rightarrow a_\mu '(x)=a_\mu (x) - \partial _\mu \rho (x)
\label{eq:13}
\en
do not change the tensor field $T_{\mu \nu }$. In the functional integral
(\ref{eq:10}) we might replace the integration over the tensor field $T_{\mu
\nu }$
by an integration over $C_\mu $ and $a_\mu $ provided that one fixes
the gauge transformations in (\ref{eq:13}). These gauge fixings can be
easily implemented and are not explicitly presented in the following.
Inserting the decomposition (\ref{eq:12}) into (\ref{eq:10}), one observes
that the $\delta $-functional constraint only acts on the fields $a_\mu $.
Integrating out $a_\mu $ one obtains
\bea
Z[j] &=& \int {\cal D} \phi \; {\cal D} \phi ^{\dagger } \;
{\cal D} C_{\mu } \; \Det ^{-2} \vert \phi \vert ^2 \;
\exp \left\{ - \int d^{4}x \; \left[ {\cal L }_D
\; + \; U \left( \vert \phi \vert ^2 \right) \right] \right\} \; ,
\label{eq:14} \\
{\cal L }_D &=& \left( \partial _\mu \phi \right)^\dagger
\partial _\mu \phi \, + \, \frac{1}{ 4 \vert \phi \vert ^2 }
\left( \partial _{\nu } F_{\nu \mu }[C] -i j_\mu - i J_\mu \right)^2
\, + \, \frac{ e^{2} }{4} F_{\mu \nu }[C] F_{\mu \nu }[C] \; .
\label{eq:15}
\ena
This is the dual formulation of scalar QED with fermions and one of our
main results. The crucial observation is that after the rescaling
$C_\mu \rightarrow C_\mu /e $ the interaction between the fermions
and the dual gauge fields is of order $1/e$.

In the dual formulation of scalar QED (\ref{eq:14}) a local functional
determinant, i.e.\ $\Det ^{-2} \vert \phi \vert ^2$, appears.
This factor could be interpreted as a functional integral measure
for $\phi (x)$ and eventually absorbed by a change of field
variables. Defining $\psi (x) = 1/ \phi (x)$, one observes that
\be
\int {\cal D} \phi ^{\dagger } \; {\cal D } \phi \;
\Det ^{-2} \vert \phi \vert ^2 \; \exp \{ - \int d^4x \; \vert
\partial _\mu \phi (x) \vert ^2 + \cdots \} \; = \;
\label{eq:18}
\en
$$
\int {\cal D} \psi ^{\dagger } \; {\cal D } \psi \;
\exp \{ - \int d^4x \; \vert \frac{ \partial _\mu \psi (x) }{
\psi ^2 (x) } \vert ^2 + \cdots \} \; .
$$
The field $\psi (x)$ has a flat measure and a non-polynominal kinetic
term.

In the following, we will consider the Higgs phase of the model
where the scale symmetry
is either broken at classical level by a specific choice of the potential
$U$ or at quantum level by a scale anomaly of the scalar sector.
In any case, the vacuum expectation value of the scalar condensate
is non-zero, i.e.\
$v^2/2:= \langle \phi ^\dagger \phi \rangle \not= 0$.
In the familiar interpretation of the Higgs phase, the
renormalisation procedure is constructed with a renormalised
tadpole parameter $v$ fixing the physical scale of the theory.
Since in the Lagrangian $L_D$ the parameter $v$ arises both as a
scale for the higher dimensions interactions and as a coupling
constant for the four fermion interaction, we will need to look
for an alternative renormalisation procedure (see below). In particular,
we will further constrain the scalar sector in order to obtain
a non-trivial strong coupling expansion later on. For this purpose,
we decompose $v^2/2$ in a part solely stemming from the scalar sector
$(\langle \phi ^\dagger \phi \rangle _0)$ and in a part due to
radiative corrections from photons and fermions, i.e.
\be
\frac{1}{2} v^2 \; := \; \langle \phi ^\dagger \phi \rangle \; = \;
\langle \phi ^\dagger \phi \rangle _0
\; + \; {\cal V} (e^2) \; .
\label{eq:18b}
\en
Our additional definition is that the parameters of the scalar sector
are chosen in order that $\langle \phi ^\dagger \phi \rangle _0$
in (\ref{eq:18b}) gives the main contribution to
$\langle \phi ^\dagger \phi \rangle $.
In this case, $v^2$ is independent of the coupling
strength $e$.  This is the crucial constraint to the scalar sector
which allows to define an expansion in powers of $1/e^2$.

To appreciate this point, we consider for a moment the alternative
case, where ${\cal V }(e^2)$ must be retained in (\ref{eq:18b}).
In order to be specific,
we choose that $v^2 = \hbox{const.}\, /e^2$. This choice is ad hoc and
only serves to illuminate our point. Introducing a photon mass parameter
by defining $m_{\gamma } = ev$, our
choice corresponds to $m_\gamma $ independent of $e$. In that case,
the strong coupling series can
be rearranged to yield the standard perturbation theory
expansion in positive powers of $e$. From this result,
we conjecture that the strong-coupling expansion provides no further
information of the Higgs phase of QED, if
$\langle \phi ^\dagger \phi \rangle $ is not mainly dominated by the
scalar sector.

One might question whether the contributions from scalar fields
$\phi (x) \approx 0 $ (although they are suppressed in the Higgs phase)
are well defined in (\ref{eq:14}). We note that in this limit the
dangerous term at the right-hand side of (\ref{eq:15}) becomes
a functional $\delta $-function, i.e.
\be
\lim _{\phi (x) \to 0} \Det ^{-2} \vert \phi \vert ^2 \;
\exp \{ - \int d^4x \; \frac{1}{ 4 \vert \phi \vert ^2 }
\left( \partial _{\nu } F_{\nu \mu }[C] -i j_\mu - i J_\mu \right)^2 \}
\label{eq:18a}
\en
\be
\rightarrow \;
\delta ^{(4)} \left( \partial _{\nu } F_{\nu \mu }[C] -i j_\mu -
i J_\mu \right) \; ,
\nonumber
\en
where the functional $\delta $-function extends over $\mu =1 \ldots 4.$
Obviously, the contributions from configurations
$\phi (x) \approx 0$ are well defined. We finally note that
(\ref{eq:18a}) might be the starting point to study the Coulomb phase
of the model. We do not follow this line here, since we do not
expect to gain further informations about this phase. The basic
idea of this paper is to choose the scalar sector appropriately
in order to make strong coupling effects transparent.

\subsection{ Symmetries and gauge fixing }

The theory described by (\ref{eq:14},\ref{eq:15}) is
invariant under two local U(1) transformations, one corresponding
to the U(1) gauge invariance (\ref{eq:4}) mediated by the photon field,
the other is the dual gauge invariance (\ref{eq:13}) introduced to
reduce the number of artificial degrees of freedom.
It is instructive to check explicitly these invariances. For this purpose
we decompose the complex scalar field, i.e.
\be
\phi (x) \; = \; \rho (x) \, \exp \{ i \pi (x) \} \; ,
\label{eq:25}
\en
rearrange the dual Lagrangian in (\ref{eq:15}), i.e.
\bea
{\cal L }_D &=&
\partial _\mu \rho \partial _\mu \rho
\, + \, U \left( \rho ^2 \right) \; + \;
\frac{1}{4} F_{\mu \nu } F_{\mu \nu } \; + \; \bar{q} (i \dslash +im) q
\, - \, \partial _\mu \pi \, j_\mu
\, - \, \frac{1}{ 4 \rho ^2 } j_\mu j_\mu \;
\label{eq:16} \\
&-& \frac{i}{2 e \rho ^2 } \partial _\nu F_{\nu \mu } j_\mu  \; + \;
\frac{1}{4 e^2 \rho ^2 } \partial _\nu F_{\mu \nu } \partial _\sigma
F_{\mu \sigma } \; ,
\nonumber
\ena
where we have also added the fermion kinetic term. We specialise from
here on to a current $j_\mu (x)$ associated to Dirac fermion matter
fields of mass $m$, denoted $q(x)$, so that
$j_\mu = \bar{q} \gamma _\mu q$ is the fermionic current.
The last two terms of this
equation contain the interactions which become weak in the strong
coupling limit.
Using the definition (\ref{eq:7}), one finds that the $\pi $-field
transforms under gauge rotations (\ref{eq:4}) as
$\pi '(x) = \pi (x) + \alpha (x) $, whereas the field $\rho (x)$
is gauge invariant. Therefore
$\partial _\mu \pi (x) $ transforms like a gauge potential
such that the fermion kinetic term and the term $\partial _\mu \pi j_\mu $
combined together are gauge invariant.
The other terms are explicitly invariant under the transformation
of the initial U(1) gauge group, spanned by $\alpha (x)$.
Since neither $F_{\mu \nu }$ nor
$j_\mu $ change under the {\it dual } gauge transformation (\ref{eq:13}),
the dual Lagrangian (\ref{eq:16}) is also gauge invariant under the
dual U(1) group.

For explicit calculations one must fix the gauge of the two
U(1) groups. In order to fix the gauge of the initial U(1) group
(\ref{eq:4}), we follow the standard procedure first proposed
by Fadeev and Popov and introduce the constant
\be
\int {\cal D} \alpha (x) \; \delta \left(
\langle \rho \rangle \pi (x) \, - \, B(x) \right) \; = \; \hbox{const.}
\label{eq:3.1}
\en
into the functional integral (\ref{eq:14}).
Since the inserted constant does not depend on $B$, one might
average over the $B$-field with a weight
\be
\exp \{ - \int d^4x \, \frac{1}{\xi } \partial _\mu B \partial _\mu B
\} \; ,
\en
where $\xi $ is a gauge parameter. The resulting part of
the dual Lagrangian (\ref{eq:16}) containing the $\pi $-field
becomes
\be
- \, \partial _\mu \pi (x) \, j_{\mu }(x) \; + \;
\frac{v^2}{2 \xi } \partial _\mu \pi (x) \partial _\mu \pi (x) \; .
\en
We now can safely integrate out the $\pi $-field and obtain an
additional current-current interaction,
\be
L_{fix} \; = \; \frac{\xi}{2v^2} \partial _\mu j_\mu \,
\partial ^{-2} \, \partial _\nu j_\nu \; .
\label{eq:3.2}
\en
We have chosen this particular class of gauges, since the
renormalisation of the strong coupling limit of the model becomes
transparent.
An analogous situation holds for the standard model of
the electro-weak interactions with $R_\xi $-gauge fixing~\cite{ch84}.
For a particular choice of the gauge parameter the model becomes
renormalisable by power counting, whereas for a different choice
the content of physical particles is obvious and renormalisability
is hidden.

In order to fix the gauge of the {\it dual } U(1) group, we
employ the gauge condition $\partial _\mu C(x) =0$, which
can be easily implemented by adding
\be
L_{fix}^D \; = \; \frac{1}{2 \xi ^D} \left( \partial _\mu
C_\mu (x) \right)^2
\label{eq:3.3}
\en
to the Lagrangian (\ref{eq:16}).

\subsection{ Strong coupling fermion interaction }

It is now straightforward to calculate the fermion current current
interaction by using\footnote{ We do
not use the notion of a ''photon propagator'' for $\Delta _{\mu \nu }$
in order to avoid confusion, since photons were integrated out
exactly in the last section.}
\be
\exp \{ \frac{1}{2} \int d^4x \; d^4 y \; j_\mu (x) \Delta _{\mu \nu }
(x-y) j_\nu (y) \} \; = \;
\int {\cal D }C \; e^{- \int d^4x \; L_{D}[j] } \; ,
\en
\be
\Delta _{\mu \nu }(x-y) \; = \; \frac{ \delta ^2 \, \ln Z[j] }{
\delta j_\mu (x) \, \delta j_\nu (y) } \; .
\label{eq:19}
\en
The Feynman rules are easily
extracted from the Lagrangian (\ref{eq:16}). The contributions to the
four fermion interaction are presented in figure \ref{fig:1}.
Since the fermion contact interaction is not suppressed by a factor
$1/e$, we must sum up these diagrams (zeroth order diagrams in
figure \ref{fig:1}). The net-effect of this resummation is that
the fermion contact interaction acquires a new strength, i.e.
\be
\frac{1}{v^2} \gamma _\mu \times \gamma _\nu \rightarrow
G_0(\Lambda ) \gamma _\mu \times \gamma _\nu \; ,
\label{eq:19a}
\en
where $G_0$ depends on the ultraviolet cut-off $\Lambda $.
The function $G_0(\Lambda )$
can be easily obtained by a direct resummation of the zeroth order
graphs in figure \ref{fig:1}. Since we will
go beyond the strong coupling expansion (we will perform a resummation
consistent with the strong coupling expansion) in order to discuss
the spontaneous symmetry breakdown in the next section, there is no
need to present the explicit form of $G_0(\Lambda )$ in the strong
coupling approach.

In next to leading order, one must take into account contributions
from the dual gauge field. One obtains
\be
\Delta _{\mu \nu }(k) \; = \; G_0(\Lambda )
P ^\xi _{\mu \nu } \; - \; \frac{1}{e^2 v^4} \int d^4x \; e^{-ikx} \;
\left\langle
T \, \partial _\sigma F_{\sigma \mu }[C](x) \,
\partial _\rho F_{\rho \nu }[C](0) \, \right\rangle \; ,
\label{eq:20}
\en
where $P^\xi _{\mu \nu } = \delta _{\mu \nu } - \xi \hat{k}_\mu \hat{k}_\nu $
with $\hat{k}_\mu = k_\mu /\sqrt{k^2}$.
The first term in (\ref{eq:20}) arises from the fermion contact
interaction and from the gauge fixing part (\ref{eq:3.2}). The second term
is correct to all orders and only limited by the order of the dual
photon propagator. In the following, we will neglect fermion
polarisation effects to the dual photon propagator (quenched
approximation). The fermion loop effects are of order $1/e^4$ implying
that they can be disregarded at the level $1/e^2$.
Using the propagator of the dual photon (to all orders and without
fermion polarisation effects), i.e.
\be
\Delta ^{(d)} _{\mu \nu } (k) \; = \; P _{\mu \nu }
\frac{1}{ k^2 \, + \, k^4/ e^2 v^2 } \; + \; \frac{ \xi ^D }{ k^2 }
\hat{k}_\mu \hat{k}_\nu \; , \; \; \; P_{\mu \nu } =
P_{\mu \nu }^{ (\xi=1) } \; ,
\label{eq:21}
\en
a direct calculation of (\ref{eq:20}) yields the correlation function
consistent with the order $1/e^2$ of the strong coupling expansion
\be
\Delta ^{(2)} _{\sigma \nu }(k) \; = \;
\left( G_0(\Lambda ) - \frac{1}{v^2} \right) \delta _{\mu \nu }
\; - \; \left( G_0(\Lambda ) \xi - \frac{1}{v^2} \right) \hat{k}_\mu
\hat{k}_\nu \; + \; P_{\mu \nu } \frac{ e^2 }{ k^2 + e^2 v^2 } \; .
\label{eq:23}
\en
The components $-1/v^2$ in $A$ and $B$ arise from reexpressing
the part from the dual photon in (\ref{eq:20}).
The main observation is that the strong coupling fermion interaction
coincides with the interaction mediated by massive photons
supplemented by additional Nambu-Jona-Lasinio interactions given by
the first two terms in (\ref{eq:23}).

\section{ Spontaneous chiral symmetry breaking  }
\label{sec:3}

The last term of the fermion interaction (\ref{eq:23}) in next
to leading order of the strong coupling expansion can be interpreted
in terms of an exchange of photon with mass $ev$.
It is encouraging that the strong coupling approach reproduces the
effects which seemingly stem from a massive photon. Indeed these effects
are also observed in lattice scalar QED in the Higgs phase~\cite{ev87}.
Chiral symmetry breaking within massive
QED (without the modifications by a Nambu-Jona-Lasinio contact
interaction) has been studied previously by Kondo~\cite{ko89} in the
zero charge limit $v/\Lambda = const.$ and in the case
$v/\Lambda \rightarrow 0$.
In the following, we will consider the general case of a fixed
value for $\rho = 4 \pi ^2 v^2/ 3 \Lambda ^2 $.
Here we have defined our model by the scalar sector being strong
implying by definition that the fermion and gauge field contributions
to $v^2$ are negligible. This leads to a $e$ independence of $v$ (and
therefore of $\rho $).
Our considerations differ from those of Kondo~\cite{ko89} by including
the four fermion contact interaction and from those of
Bardeen~\cite{bar86} by taking into account the mass of the photon.

The Dyson-Schwinger equation in Euclidean space
for the fermion propagator $S(p)$
consistent with the next to leading order of the strong coupling
expansion is
\be
-i \pslash \, + \, m \, + \, \int \frac{ d^4k }{ (2\pi )^4 } \;
\gamma _\mu \, \Delta ^{(2)} _{\mu \nu }(p-k) \, S(k) \, \gamma _\nu
\; = \; S^{-1}(p) \; ,
\label{eq:2.1}
\en
where $\Delta ^{(2)}_{\mu \nu }(k)$ is the strong coupling fermionic
correlation function (\ref{eq:23}).
Using the ansatz
\be
S(p) \; = \; \frac{i}{ Z(p) \pslash \, + \, i \Sigma (p) } \; ,
\label{eq:2.2}
\en
equation (\ref{eq:2.1}) yields two equations for the
scalar functions $Z(p)$ and $\Sigma (p)$. A simplification, which
is known to be reliable for this application~\cite{ko89}, and whose
virtue is to allow an analytical treatment,
is achieved by replacing the $\Delta ^{(2)} _{\mu \nu }$
in (\ref{eq:2.1}) by
\be
\Delta ^{(2)} _{\mu \nu }(p-k) \; = \; A
\delta _{\mu \nu } \; - \; B
(\hat{p}-\hat{k})_\mu (\hat{p}-\hat{k})_\nu \; + \; P_{\mu \nu }
\frac{ e^2 }{ (p-k)^2 } \frac{ 1 }{ 1 + e^2 v^2/ \hbox{max} (p^2,k^2) }
\; ,
\label{eq:2.3}
\en
where $A=G_0 - \frac{1}{v^2} $ and $B=G_0 \xi - \frac{1}{v^2} $.
Using (\ref{eq:2.3}) and analytic continuation to Euclidean space,
a straightforward calculation yields
\bea
\Sigma (p) &=& m + \int \frac{ d^4 k}{ (2\pi )^4 } \;
\frac{ \Sigma (k) }{ Z^2 k^2 + \Sigma ^2 } \,
\left( 4A-B + \frac{3e^2}{(p-k)^2} \frac{1}{1 + e^2 v^2 /
\hbox{max}(k^2,p^2)} \right)
\label{eq:2.4} \\
Z(p) &=& 1 - \frac{1}{4 p^2 } \int \frac{ d^4 k}{ (2\pi )^4 } \;
\frac{ Z (k) }{ Z^2 k^2 + \Sigma ^2} \, \tr \{
-2 A \kslash \pslash - B \pslash (\hat{\pslash }- \hat{\kslash })
\kslash (\hat{ \pslash } - \hat{\kslash } ) \} \; .
\label{eq:2.5}
\ena
We have skipped in (\ref{eq:2.5}) the term proportional $P_{\mu \nu }$
of (\ref{eq:2.3}), since this term will be zero after performing the angle
integration. This is due to the fact that this acts like the term
induced by a massless photon exchange, the
angle integration of which is known to yield zero~\cite{co89}.
We first focus on $Z(p)$ in (\ref{eq:2.5}). After the angle integration
this term is
\bea
Z(p) &=& 1 + \frac{B}{4 \pi ^2 } \left\{ \int _0 ^{p} dk \; \frac{
k^5 ( k^2- 3 p^2)}{p^4} \frac{ Z (k) }{ Z^2 k^2 + \Sigma ^2}
\right. \label{eq:2.6} \\
&+& \left. \int _p ^{\infty } dk \;
k ( p^2- 3 k^2) \frac{ Z(k) }{ Z^2 k^2 + \Sigma ^2} \; \right\} \; .
\nonumber
\ena
The considerations are further greatly simplified by choosing the gauge
\be
\xi \; = \; \frac{1}{ G_0(\Lambda ) \, v^2 } \; .
\; \label{eq:2.7}
\en
We then have $B=0$, and hence $Z(p^2)=1$ so that
no wave function renormalisation is requested.
We should note at this point that the result (\ref{eq:2.6}) holds
only in the particular separable approximation used in (\ref{eq:2.3}).
Performing the angle integration  in (\ref{eq:2.4}), the equation for
the self-energy $\Sigma (p)$ becomes
\bea
\Sigma (p) &=& \; \mu _0 + \frac{ r }{2 } \frac{1}{p^2+e^2v^2}
\int _{0} ^{p} dk \; k^3 \frac{ \Sigma (k) }{ k^2 + \Sigma ^2} \; ,
\label{eq:2.8} \\
&+& \frac{ r }{ 2 } \int _{p}^{\infty } dk \; k^3
\frac{ 1 }{k^2 + e^2 v^2} \frac{ \Sigma (k) }{ k^2 + \Sigma ^2 } \; ,
\nonumber
\ena
where $r=\frac{ 3 e^2 }{4 \pi ^2} $ and where we have introduced
\be
\mu _0 \; = \; m \; + \; 4A \int \frac{ d^4 k}{ (2\pi )^4 }
\; \frac{ \Sigma (k) }{ k^2 + \Sigma ^2 } \; .
\label{eq:2.10}
\en
The only effect of the contact interaction is to induce a shift in
the bare fermion mass, i.e.
$m \rightarrow \mu _0$. We reformulate the equation (\ref{eq:2.8})
for the self-energy in terms of a differential equation by standard
techniques (see e.g.~\cite{co89}). Taking the derivative of (\ref{eq:2.8})
with respect to $p^2$, one obtains
\be
\Sigma ' (p^2)  \; = \; - \frac{r}{4} \frac{1}{(p^2 + e^2 v^2)^2} \,
\int _{0}^{p^2} dk^2 \; k^2 \frac{ \Sigma (k) }{ k^2 + \Sigma ^2 (k) } \; ,
\label{eq:2.12}
\en
where the prime denotes the derivative with respect to $p^2$.
This equation provides the boundary condition
\be
\Sigma '(p^2=0) = 0 \; .
\label{eq:2.14}
\en
Multiplying both sides of (\ref{eq:2.12}) with $(p^2+e^2v^2)^2$ and
differentiating this equation again with
respect to $p^2$, the desired differential equation is
\be
\left[ (p^2+e^2 v^2)^2 \, \Sigma ' \right] ' \; + \;
\frac{r}{4} \frac{ p^2 \Sigma }{ p^2 + \Sigma ^2 } \; = \; 0 \; .
\label{eq:2.13}
\en
Obviously $\Sigma (p^2) \equiv 0$ is always a solution of this
equation, which in addition satisfies the boundary condition
(\ref{eq:2.14}).
Following~\cite{ko89}, we will discuss the occurrence of spontaneous
symmetry breaking by means of the bifurcation method~\cite{atk87}.
The basic idea is that
near some critical coupling $r_c$ the gap equation (\ref{eq:2.13})
allows for a second solution $\Sigma $ aside from the trivial one
$\Sigma _0 =0$ which can be represented by
\be
\Sigma \; = \; \Sigma _0 \; + \; \sigma
\label{eq:2.23}
\en
with $\sigma $ small. This implies that near the critical
coupling $\sigma $ satisfies the linearised gap-equation. One
observes that $\sigma $ does only depend on $Q:= p^2+e^2 v^2$ via
the so-called bifurcation equation
\be
\sigma ''(Q) \, + \, \frac{2}{Q} \sigma '(Q) \, + \, \frac{r}{4Q^2}
\sigma (Q) \; = \; 0 \; .
\label{eq:2.25}
\en
The general solution of this equation was e.g.\ presented in
Kondo's paper~\cite{ko89}. For completeness, we mention that
the ansatz
\be
\sigma (Q) \; = \; g \left( \ln \frac{Q}{\mu ^2} \right) / \sqrt{Q}
\label{eq:2.26}
\en
transforms the differential equation (\ref{eq:2.25}) into
\be
g''(x) \; + \; \frac{r-1}{4} \, g(x) \; = \; 0 \; ,
\label{eq:2.27}
\en
from which the solutions can be easily obtained, i.e.
\be
\vbox{
\begin{tabular}{lll}
$ \sigma (Q) \; = \; a/Q^{(1+w)/2} + b/Q^{(1-w)/2} \; ,
\; $
& $ w= \sqrt{1-r} \; , \hbo $
& $ (r < 1) $ \\
$ \sigma (Q) \; = \; c \ln \frac{Q}{d} / \sqrt{Q}  \; ,
\; $
& & $ (r = 1) $ \\
$ \sigma (Q) \; = \; e \sin \left( \omega \ln
\frac{ Q }{f} \right) / \sqrt{Q} \; ,
\; $
& $ \omega = \frac{1}{2} \sqrt{r-1} \; , \hbo $
& $ (r > 1) $ \\
\end{tabular}
}
\label{eq:2.28}
\en
where $a,b,c,d,e,f$ are integration constants.
The basic question concerning the spontaneous symmetry breaking is whether
the non-trivial bifurcation solutions (\ref{eq:2.28}) ($\sigma \not=0$)
are compatible with the boundary condition (\ref{eq:2.14}) and
the scaling limit. Since all authors~\cite{mir85,bar86,co89,ko89,kon91}
agree that the spontaneous breakdown of chiral symmetry occurs
for $r \ge 1$, we here only address the subtle question whether
it also occurs for $r<1$ in our model due to the support of the
contact interaction.
Mass renormalisation plays the central role in discussing the scaling
limit of the model. In order to address this issue, we first note that
the bifurcation solution satisfies the linearised integral equation
(\ref{eq:2.8}), i.e.
\be
\sigma (Q) \; = \; \mu _0 + \frac{ r }{4Q}
\int _{0} ^{Q} dQ' \; \sigma (Q') \; + \;
\frac{ r }{ 4 } \int _{Q}^{\Lambda ^2 + e^2 v^2 } dQ' \;
\frac{ \sigma (Q') }{ Q' } \; ,
\label{eq:2.15}
\en
where we have introduced the UV-cutoff $\Lambda $. From this equation
we deduce
\be
\sigma '(Q) \; = \; - \frac{r}{4Q^2} \int _{0}^{Q}
dQ' \; \sigma (Q') \; ,
\en
and combining this equation with (\ref{eq:2.15})
\be
\sigma (Q) \, + \, Q \sigma '(Q) \; = \; \mu _0 \, + \,
\frac{r}{4} \int _{Q}^{\Lambda ^2+e^2v^2 }
dQ' \; \frac{ \sigma (Q') }{Q'} \; .
\label{eq:2.16}
\en
The boundary condition (\ref{eq:2.14}) transforms into a constraint
for the bifurcation solution, i.e.
\be
\sigma '(Q=e^2v^2) \; = \; 0 \; .
\label{eq:2.17}
\en
For $r<1$ this condition provides a relation for the coefficients
$a,b$ in (\ref{eq:2.28}), i.e.
\be
a \; = \; \frac{ w-1 }{w+1} \, (ev)^{2w} \, b \; .
\label{eq:2.18}
\en
In order to extract the cutoff dependence of $\mu _0$ (see~\cite{co89}),
we plug the bifurcation solution (\ref{eq:2.28}) into (\ref{eq:2.16}).
One observes that the $Q$-dependence cancels, and we are left with
\be
\mu _0 \; = \; \frac{r}{2} \left[ \frac{a}{1+w}
\left( \frac{ 1  }{ \Lambda ^2 +e^2v^2 } \right)^{\frac{1+w}{2} }
\, + \, \frac{b}{1-w} \left(
\frac{ 1  }{ \Lambda ^2 +e^2v^2} \right)^{\frac{1-w}{2} } \right] \; .
\label{eq:2.19}
\en
In order to determine the scaling dimension of the current mass $m$, we
exploit the relation of $\mu _0$ to $m$ in (\ref{eq:2.10}), which
becomes
\be
\mu _0 \; = \; m \; + \; \frac{A}{4 \pi ^2} \int _0^{\Lambda ^2 }
du \; u \; \frac{ \Sigma (u) }{ u + \Sigma ^2 (u) } \; .
\label{eq:2.19a}
\en
The integral in
(\ref{eq:2.19a}) can be easily evaluated by using the differential
equation (\ref{eq:2.13}), i.e.
\be
\mu _0 \; = \; m \; - \; \frac{ A }{r \pi ^2 } \,
(\Lambda ^2 + e^2 v^2 )^2 \, \Sigma '(\Lambda ^2) \; .
\label{eq:2.20}
\en
Asymptotically the function $\Sigma (p^2)$ can be approximated
by the bifurcation solution, i.e. $\Sigma (p^2) \approx \sigma (p^2)$.
Inserting this asymptotic form in (\ref{eq:2.20}) and combining this
equation with (\ref{eq:2.19}) and (\ref{eq:2.18}), we finally
obtain
\bea
m &=& \left[ - \left( \frac{r}{1+w} - (1+w)
\frac{A \Lambda ^2 (1+ r \rho ) }{ r \pi ^2 } \right)
\frac{ (r \rho ) ^{w} }{ (1+r\rho )^w } \frac{1-w}{1+w}
\label{eq:2.21} \right. \\
&+& \left.
\frac{r}{1-w}- (1-w) \frac{A \Lambda ^2 (1+r\rho ) }{r \pi ^2 } \right]
\; \frac{ b }{ 2 \Lambda ^{1-w} (1+r\rho )^{(1-w)/2} } \; ,
\ena
where $\rho = 4\pi ^2 v^2 / 3 \Lambda ^2$. The continuum limit
can be defined provided $\rho $ and $A\Lambda ^2 = ( G_0(\Lambda )
-1/v^2) \Lambda ^2 $ tend to finite limits as $\Lambda ^2
\rightarrow \infty $.
A finite value for $ev$ corresponds to  $\rho =0 $.
One observes that the bare current mass tends to zero with
increasing regulator $\Lambda $.
{}From (\ref{eq:2.21}), one might define the renormalised mass $m _R$
and might extract the scale dependence of the mass renormalisation
constant $Z_{(m)} (\Lambda )$, i.e.
\be
m _R \; = \; Z_{(m )}(\Lambda ) m \; , \hbo
Z^{(r<1)}_{(m)} (\Lambda ) \; = \; \left( \frac{ \Lambda }{ \mu } \right)
^{1-w } \; .
\label{eq:2.22}
\nonumber
\en
This definition of the renormalised mass is in agreement with
that from Cohen and Georgi~\cite{co89}, which is extracted from
the operator product expansion. The anomalous dimension of the
fermion mass is defined by
\be
\gamma _{(m)} \; = \; \frac{ \partial \ln Z_{(m)} (\Lambda )
}{ \partial \ln \Lambda } \; .
\en
{}From (\ref{eq:2.22}) we obtain the well know result (for $r \le 1$)
\be
\gamma _{(m)} \; = \; 1 \; - \; \sqrt{1-r} \; .
\en
In order to discuss the occurrence of dynamical mass scale, we
confine our considerations to the chiral limit $ m_R =0$.
In this case, the square bracket in (\ref{eq:2.21}) must vanish.
This defines the critical line in the plane of the coupling constants
$A \Lambda ^2$ and $r$ as the one-dimensional domain in which
the model approaches the scaling limit. A direct calculation yields
\be
\frac{A \Lambda ^2 }{ \pi ^2 } \; = \;
\frac{ r }{ 1 + r \rho } \,
\frac{ \frac{1-w}{1+w} \left( \frac{ r\rho }{ 1 + r\rho } \right)^w
\, - \, \frac{1+w}{1-w} }{ \left( \frac{ r\rho }{ 1 + r\rho } \right)^w
\, - \, 1 } \; , \hbo (w=\sqrt{1-r}) \; .
\label{eq:2.23a}
\en
Th critical line is presented in figure \ref{fig:2} for several
values of $\rho $. For finite $ev$ ($\rho =0$) one finds
\be
\frac{A \Lambda ^2 }{ \pi ^2 } \; = \; ( 1 \, + \, \sqrt{1-r})^2 \; .
\label{eq:2.24}
\en
Here we end up with the same conclusion Bardeen et al.\ have
presented in their early work~\cite{bar86}. The chiral
symmetry is broken even for $r<1$. A smaller value of the
gauge coupling constant $r$ can be compensated by a larger
value of the renormalised Nambu-Jona-Lasinio coupling constant
$A \Lambda ^2$. In addition, we find that for finite $\rho $
a larger strength of the contact interaction is needed to break the
chiral symmetry. This is an intuitive result, since for a large photon
mass the interaction mediated by the photon is suppressed.

One might easily extend the results to the case $r>1$ by following the
steps above. We only present the final result, i.e.
\be
\frac{A \Lambda ^2 }{ \pi ^2 } \; = \; \frac{1}{1+r \rho }
\, \left[ 1 - 4 \omega ^2 - \frac{ 4 \omega }{ \hbox{tan} \,
\omega \sigma } \right] \; , \hbo
\sigma = \ln \frac{ r \rho }{ 1 + r \rho } \; .
\label{eq:2.30}
\en
One observes that for large values of $r$ the contact interaction
approaches a constant value, i.e.
\be
\frac{A \Lambda ^2 }{ \pi ^2 } (r \rightarrow \infty ) \; = \;
\frac{ 4 \rho -1 }{\rho } \; .
\label{eq:2.31}
\en
For small values of $\rho $, i.e.
\be
\rho \; \le \; \frac{1}{r} \, \frac{ 1 }{ \exp \{
\frac{ 2 \pi }{ \sqrt{ r-1 } } \} -1 } \; ,
\en
the tangent in (\ref{eq:2.30}) develops zeros, which subsequently
give rise to a rapidly varying critical curve $A(r)$. The theory
is infrared sensitive. In contrast, for a sufficiently large photon mass
$(\rho > 0.015496 \ldots )$, the influence of the infrared regime is
diminished, and the critical curve $A(r)$ smoothly approaches
its asymptotic value for large $r$.

\section{ Conclusions }

In this paper, we have provided further support to the ideas of
strongly coupled QED as it was first raised by Miransky~\cite{mir85} and
further developed by Bardeen et al.~\cite{bar86} and
Cohen and Georgi~\cite{co89}. For this purpose, strongly coupled
scalar QED with fermions was investigated in its dual formulation
which provides a perturbative expansion with respect to the inverse
coupling constant, i.e.\ $1/e$. In the first part of the paper,
we derived the strong coupling expansion by means of functional
integral techniques. We modeled the scalar sector in order to
allow for a rigorous treatment of the fermion and gauge-field sectors.
The assumptions which define the scalar theory were first that
the contributions from fermions
and dual photons to the vacuum expectation value $v$ of the scalar field
can be neglected; secondly, the fluctuations of the scalar field around
its vacuum expectation value acquire a large mass implying that
they have decoupled from the theory. The first supposition implies
that the vacuum expectation value $v$ does not depend on the
coupling strength $e$. One might wonder whether this choice of the
scalar sector is a natural one.
In fact, we have not found any contradiction of the
assumptions above with the lattice simulations~\cite{ev87,fra79}
of scalar QED in the Higgs phase, where the scalar fields $\phi $
interact via a $\phi ^4$-interaction.

Subsequently, the fermion interaction was calculated
consistent with the second order of the strong coupling
expansion in the quenched approximation (fermion polarisation effects
to the propagation of the dual photon were neglected). We found
that the fermion interaction can be effectively described by a
Nambu-Jona-Lasinio contact interaction supplemented by the exchange
of a photon with mass $ev$. Up to this mass term, this model
was precisely proposed by Bardeen et al.~\cite{bar86} in order to
obtain a consistent non-perturbative renormalisation of QED.
The previous considerations~\cite{mir85,bar86,co89,ko89,kon91} are based
on a partial resummation of perturbation theory with respect to
the coupling strength $e$. Here we justify the approach from a
strong coupling point of view implying that the previous results
might be also valid for the range of coupling constants of interest
($e^2 \gg 1$).

Our model mainly differs from the model considered by Bardeen et
al.~\cite{bar86} by the presence of a photon mass
$ev$. Spontaneous symmetry breaking in massive QED
(without the Nambu-Jona-Lasinio four fermion interaction)
was previously studied by Kondo~\cite{ko89} in the zero charge limit
(the photon mass scales with the cutoff) and beyond (the photon mass
is finite). We extended this work by
including the fermion contact interaction as it naturally arises
in the strong coupling approach. We found that chiral symmetry
breaking also occurs for small values of the gauge coupling
constant due to the presence of the contact interaction.
This is in qualitative agreement with the case of a massless photon
studied by Bardeen et al.

\bigskip
{\bf Acknowledgements: }
Part of this work was done while one of the authors (KL) was participating
in the 1995 Spring program on ``Chiral dynamics in hadrons and nuclei"
at the Institute for Nuclear Theory, University
of Washington, Seattle. KL would like to thank the participants for
helpful discussions and the INT for the hospitality.

\newpage
\centerline{ \bf \large Figure captions }
\vspace{ 2cm }
\begin{figure}[h]
\caption{ Diagrams contributing to the four fermion interaction
  within the strong coupling expansion. Wavy line: dual photon
  propagator; solid line: fermion propagator; crossed circle:
  $1/e^2$-correction to the dual photon propagator. }
\label{fig:1}
\end{figure}
\begin{figure}[h]
\caption{ The critical line separating the phase with spontaneous
   chiral symmetry breaking (SSB) from the symmetry restored phase (RS)
   in the plane of coupling constants $A \Lambda ^2/\pi^2 $ and
   $r=3 e^2 /4 \pi ^2$ for several values of $\rho = 4 \pi ^2
   v^2 / 3 \Lambda ^2 $. }
\label{fig:2}
\end{figure}
\end{document}